\documentclass{aa}

\usepackage[dvips]{color}
\usepackage{ulem}

\usepackage{graphicx}
\usepackage{txfonts}

\usepackage{subfigure}

\begin{document}

\title {On the RZ Draconis Sub-stellar Circumbinary Companions}

\subtitle{Stability Study of the Proposed Sub-stellar Circumbinary System}

   \author{Tobias C. Hinse\inst{1}
          \and
          Jonathan Horner\inst{2,3}
	  \and
	  Jae Woo Lee\inst{1}
	  \and
	  Robert A. Wittenmyer\inst{3,4}
	  \and
	  Chung-Uk Lee\inst{1}
	  \and
	  Jang-Ho Park\inst{1}
	  \and
	  Jonathan P. Marshall\inst{4}
	  }

   \institute{Korea Astronomy and Space Science Institute,
              776 Daedukdae-ro, Yuseong-gu, 305-348 Daejeon, 
	      Republic of Korea. \\
              \and
	      Computational Engineering and Science Research Centre, 
	      University of Southern Queensland, Toowoomba, Queensland 4350,                   Australia.\\
	      \and
	      Australian Centre for Astrobiology, University of New South 
	      Wales, Sydney 2052, Australia. \\
              \and
	      School of Physics, University of New South Wales, Sydney 2052,
	      Australia. \\
             }

\date{Received:---------; Accepted:----------}

\abstract
{Recently, using the light-travel time effect, planets and sub-stellar 
companions have been proposed to orbit around binary star systems (aka circumbinary companions) as a result of variations in timing of observed eclipses. For the majority of these systems the proposed orbital architecture features crossing orbital configurations as a result of high eccentricities for one, or both, of the companions. For such systems, strong mutual gravitational interactions are expected, resulting in catastrophic orbital instabilities, or collisions between the proposed components, on very short timescales.}
{In this paper, we re-examine the primary and secondary eclipse timings of the short-period and semi-detached binary RZ~Draconis (RZ~Dra), as originally presented in Yang et al. (2010). In their work, the proposed companions have masses of around $\simeq 0.07$ and $\simeq 0.18~M_{\odot}$ with the inner companion on an orbit with moderate eccentricity (0.46) having its apocenter distance crossing the orbit of the outer companion. We show that the companions proposed by Yang et al. (2010) follow highly unstable orbits. In an attempt to find a stable system we have searched the underlying $\chi^2$ parameter space for a best-fit model and carried out an orbital stability study in order to 
test possible best-fit models. If the binary period changes are truly due to additional massive companions in a hierarchical configuration, they must follow stable orbits.}
{For numerical orbital stability calculations we use well-established orbit integration routines. Computations were carried out using large-scale 
multi-CPU computing environment. Data analysis of times of primary and secondary eclipse is based on the Levenberg-Marquardt least-squares minimisation algorithm using the two-body Keplerian light-travel time effect model.}
{Despite the wide variety of potential models tested for the RZ~Dra system in this work, we found very few models that were stable for even one million years, with the vast majority of systems tested falling apart on timescales of just hundreds of years. It seems likely therefore, that the observed timing variations are not solely the result of massive, unseen companions.}
{}

{}

\maketitle

\section{Introduction}

A hierarchical \citep{Evans1968} multi-body star system is believed to be formed through one or more formation channels. First, \cite{vandenBerkEtAl2007} considers interaction/capture mechanisms during the formation and dynamical evolution of a globular star cluster. A second mechanism that could explain the existence of such systems is that they could be formed directly from a massive primordial disk involving accretion processes and/or local disk instabilities \citep{LimTakakuwa2006,DucheneEtAl2007,MarzariEtAl2009}. A third mechanism follows a chaotic erosion process of a non-hierarchical star system by angular momentum and energy exchange via mutual gravitational interactions. In the latter case, and considering an initial triple-system, \cite{Reipurth2000} provides a schematic outline of three stages that could produce a close binary system with a circumbinary disk from redistribution of circumstellar material after chaotic interactions. The formation of the tightly bound central binary is followed by the transport of the third member to a wider orbit as a result of conservation of energy. In extreme cases, this can result in the third member being ejected completely from the system leaving a tightly-packed close binary on a quasi-Keplerian orbit.

A particular example of a hierarchical multi-body system is a so-called circumbinary system (aka companions on P-type orbits \citep{Schwarz2011}) in which one or more massive objects orbit a binary star system. Such systems has been recently discovered by {\sc Kepler} and the {\sc Planet Hunters} community\footnote{www.planethunters.org} \citep{DoyleEtAl2011,WelshEtAl2012,OroszEtAl2012a,OroszEtAl2012b}. \cite{Ofir2009} presents the results for a search of circumbinary companions based on {\sc CoRoT} data. The planet orbiting the binary PH-1 \citep{SchwambEtAl2013, KostovEtAl2013} is a particularly exotic example of such a system. Here the binary is a member of a quadruple (or quaternary) hierarchical system where two binary pairs form a gravitationally bound star system. Similar in nature though with no evidence of planetary companions is the HD98800 quadruple system \citep{FurlanEtAl2007}. Other types of hierarichical star systems reside in so-called S-type \citep{Schwarz2011} configurations where one body is orbiting one component of a binary pair. Several examples of such systems have been reported in the literature \citep{Neuhauser2007,Chauvin2007}.

A {\bf well-known} technique to detect a hierarchical circumbinary systems is to measure and monitor timing variations of the mid-eclipse times of the central binary (aka times of minimum light). {\bf For a detailed description of its application on detecting circumbinary companions of planetary mass we refer the interested reader to \citet{DeegEtAl2000} and \citet{DoyleDeeg2004}. This technique has recently begun to be applied to the excellent timing data collected by the {\sc Kepler} mission, resulting in the recent announcement of the first sub-stellar mass circumbinary companion discovered from that data, orbiting KIC002856960 \cite{LeeEtAl2013a}.} The fundamental principle of the light-travel time effect (LTT\footnote{sometimes also referred to as LITE \citep{BoHe1996}}) makes use of the motion of the binary around the total barycenter of the system. Due to the finite speed of light, the eclipses exhibit delays or advances in the timings of minimum light depending on the orbital position of the binary relative to the observer \citep{Irwin1952,Irwin1959}. This method is particularly attractive as it is observationally time-effective involving only photometric CCD measurements.

In recent times, single and multi-body sub-stellar circumbinary companions to known eclipsing binary systems have been proposed using ground-based timing measurements \citep{LeeEtAl2011,LeeEtAl2012,LeeEtAl2013a,LeeEtAl2013b}. The 
same technique 
was used to detect candidate circumbinary companions of planetary nature: CM Draconis \cite[CM~Dra, one companion]{DeegEtAl2000}, DP Leonis \cite[DP~Leo, 
one companion]{QianEtAl2010a, BeuermannEtAl2011}, HW Virginis \cite[HW~Vir, 
two companions]{LeeEtAl2009a}, NN Serpentis \cite[NN~Ser, two companions] {QianEtAl2009, BeuermannEtAl2010, BeuermannEtAl2013}, UZ Fornazis 
\cite[UZ~For, two companions]{PotterEtAl2011} and HU Aquarii \cite[HU~Aqr, 
two companions]{QianEtAl2011}, QS Virginis \cite[QS~Vir, one or two 
companions]{QianEtAl2010b, Almeida2011}. Recently additional circumbinary companions were proposed utilising the LTT effect: RR Caenis \cite[RR~Cae, one companion]{Qian2012a}, NSVS 14256825 \cite[two companions]{Almeida2013} and 
NY Virginis \cite[NY~Vir, one companion]{Qian2012b}.

However, the existence of the proposed multi-body systems has been cast in doubt as a result of a number of studies of the dynamical stability of their orbital architectures. The proposed companions around HU~Aqr have been studied in detail, and a series of studies have revealed them to be dynamically unfeasible \citep{HUAqr,HinseEtAl2012a,HUAqr2, FunkEtAl2011, FunkEtAl2012, GozdziewskiEtAl2012}. The same is true of HW~Vir \citep{HWVir} and NSVS~14256825
\citep{NSVS,HinseEtAl2014}. Indeed, of those systems studied in this way, the only one to withstand dynamical scrutiny is that around NN~Ser \citep{NNSer,BeuermannEtAl2013}, although a recent study of the evolution of the central binary in the NN~Ser system suggests that it is unlikely that planetary companions on the proposed orbits could have survived the system's post-main sequence evolution intact \citep{Mustill2013}. Furthermore, \cite{HinseEtAl2012b} showed that the two sub-stellar companions orbiting around the SZ Herculis \cite[SZ Her, two companions]{LeeEtAl2012} binary also follow highly unstable orbits. The same situation is also seen for the QS~Vir system where \cite{QSVir} could show that the proposed two-companion system is highly unstable. 

Finally, \citet{ParsonsEtAl2010} present photometric follow-up observations of a number of eclipsing post-common-envelope binaries where they have been able to rule out previous claims for single-object circumbinary companions (e.g., \citet{QianEtAl2009}, \citet{QianEtAl2010b}).

In this work we re-examine the observed timing dataset of RZ~Dra as presented by \cite{YangEtAl2010}. Those authors propose the existence of two additional low-mass dwarfs from two distinct quasi-sinusoidal variations in the times of mutual eclipses. Section 2 presents a dynamical stability analysis of the nominal orbital parameters as derived by \cite{YangEtAl2010}. We then continue and give an outline of our data analysis based on the light-travel time effect and describe the least-squares methodology in section 4, where we present a new best-fit model of the two proposed companions. A dynamical analysis of our 
new model is presented in section 5 and we finish with concluding remarks in section 6.

\section{Orbital stability of \cite{YangEtAl2010} model}

\begin{table}
\centering
\begin{tabular}{lccl}
\hline
Parameter & RZ~Dra(AB)C & RZ~Dra(AB)D & Unit \\
\hline
$m~\sin~I$ & 0.074~$\pm$~0.004 & 0.175~$\pm$~0.009 & $M_\odot$\\

$a~\sin~I$ & 12.37~$\pm$~1.23 & 23.88~$\pm$~2.63 & AU\\

$e$ & 0.46~$\pm$~0.02 & 0.287~$\pm$~0.007 & \\ 

$\omega$ & 106~$\pm$~1 & 158~$\pm$~3 & deg\\

$T$ & 2,440,800.9~$\pm$~69.0 & 2,440,309.2~$\pm$~802.6 & HJD\\

$P$ & 27.59~$\pm$~0.10 & 75.62~$\pm$~2.20 & year\\
\hline
\end{tabular}
\caption{Orbital and mass parameters of the two sub-stellar companions 
proposed to orbit the RZ~Dra Algol-type binary system, taken from \citet{YangEtAl2010} (their Table 5). Here we use the naming convention as proposed by \cite{HessmanEtAl2010} to denote eclipsing binaries with additional companions. }
\label{Yang10_params}
\end{table}

In their 2010 paper, Yang et al. proposed that the observed variations in the timings of eclipses between the components of the Algol-type binary star system RZ Dra could be explained by the presence of two massive, unseen companions moving on eccentric long-period orbits, as detailed in Table \ref{Yang10_params}. On their nominal best-fit orbits, these two proposed objects can approach one another remarkably closely - with the innermost object (RZ Dra (AB) C)\footnote{We adopt the naming convention as proposed by \cite{HessmanEtAl2010}.} having a best-fit apastron distance of 18.06 AU, and the outermost (RZ Dra (AB) D) having a nominal best-fit periastron distance of 17.02 AU. In other words, the nominal best-fit model for the two proposed companions have significant overlap - these stellar-mass companions can cross one another's orbits. 

\begin{figure*}
\sidecaption
\includegraphics[width=120mm]{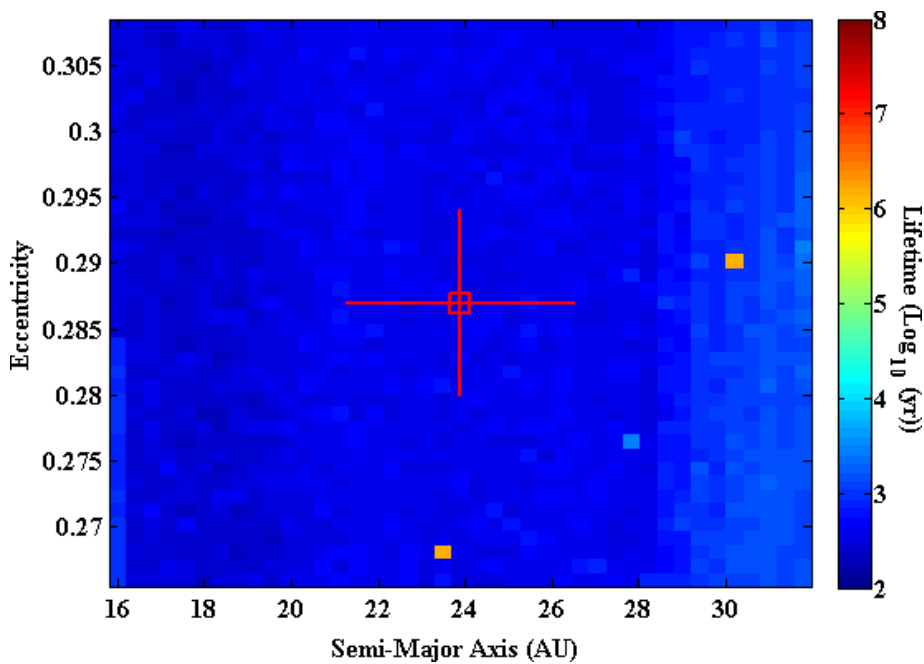}
\caption{Dynamical stability of the RZ Dra system, as a function of the semi-major axis, $a$, and eccentricity, $e$ of RZ Dra (AB) D, for the solution presented in Yang et al., 2010. The nominal best-fit orbit for RZ Dra (AB) D presented in that work is located in the centre of the red box, at $a = 23.88$ AU and $e = 0.287$. The $1\sigma$ uncertainties on the semi-major axis and eccentricity are shown by the red crosshairs radiating from the box. The lifetime at each of the 1681 $a-e$ locations plotted in this figure (41 unique $a$ and $e$ values spanning $\pm 3\sigma$ from the nominal best-fit orbit) is the result of 75 distinct simulations, spanning 5 unique values of mean anomaly, $M$, and 15 unique values of the argument of periastron, $\omega$. The mean lifetime can be seen to vary between $\simeq 100$ and $\simeq 1600$ years, demonstrating that the system as proposed in Yang et al. exhibits extreme dynamical instability. {\it See electronic version for colors}.}
\label{original}
\end{figure*}

As we have shown in previous works (e.g. \citet{HUAqr, HUAqr2, HWVir, NSVS}), such mutually encountering orbital architectures typically lead to significant dynamical instability, often on timescales of just a few hundred years. Given the age of the systems studied, such instability clearly rules out a planetary origin for the observed variations in the aforementioned systems, suggesting instead that some other astrophysical effect is the cause of the observed timing variations. It is clearly therefore important to examine the dynamical feasibility of the proposed companions to RZ Dra, in order to see whether the ``companion hypothesis'' of Yang et al. stands up to close scrutiny.

In order to study the dynamical stability of the proposed companions (RZ~Dra (AB)~C and D), we used the HYBRID integration algorithm within the $n$-body dynamics package {\sc mercury} \citep{Mercury}. In this work we used a constant integration time step of 1 day in all our orbit integrations. The 
error tolerance parameter was set to one part in $10^{12}$ which ensures accurate integrations (using the Bulirsch-Stoer algorithm) of 
high-eccentricity orbits and eventual close encounters. As in our earlier work (e.g. \citet{HR8799, HUAqr, HWVir, HD155358, HD204313, HD142}), we performed a large number of discrete simulations, each following the dynamical evolution of one potential RZ Dra system. As in those earlier works, we held the initial orbital elements of the better constrained of the two companions (in this case, RZ~Dra (AB)~C) fixed at their nominal best-fit values (as detailed in Table \ref{Yang10_params}). We then systematically varied the initial semi-major axis, $a$, eccentricity, $e$, argument of periastron, $\omega$, and mean anomaly, $M$, of the orbit of RZ~Dra (AB)~D across their full $\pm$ 3$\sigma$ range, thereby creating a wide variety of test systems, which were then integrated for a period of 100 Myr. In each test system, the dynamical evolution of the two companions was followed until a break-up of the system was detected - either through the ejection of one or other of the two companions (to a distance of 50 AU - sufficiently distant that the sub-stellar components could only reach this distance if significant mutual interaction has occurred), through mutual collision between the two companions, or when one of the companions collided with the central bodies. If the test system fell apart in this manner, the time of its demise was recorded.

In total, the dynamical evolution of 126075 trial systems was considered within our numerical stability analysis. 41 distinct values of $a$ were tested, spread evenly across the $\pm$ 3$\sigma$ range (i.e. between $a = 15.99$ and $a = 31.77$ AU). For each of these $a$ values, 41 unique values of $e$ were tested, again evenly distributed over the $\pm$ 3 $\sigma$ range (between $e = 0.266$ and $e = 0.308$). For each of these 1681 $a-e$ pairs, 15 unique $\omega$ values were tested, and at each of these 25215 $a-e-\omega$ trials, 5 unique mean anomalies were tested. 

\begin{figure}
\mbox
{
{\scalebox{0.185}{\includegraphics{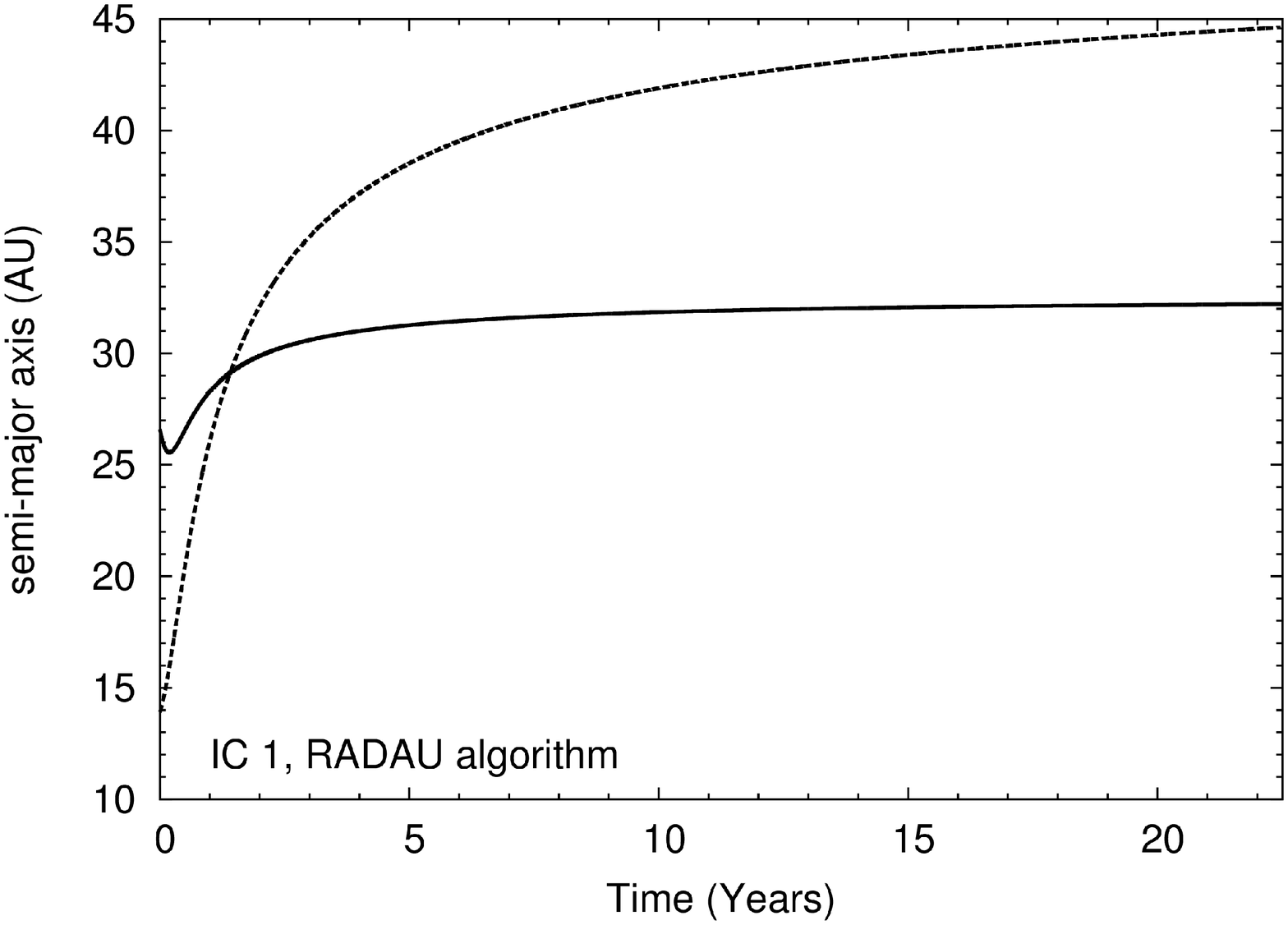}}}
{\scalebox{0.185}{\includegraphics{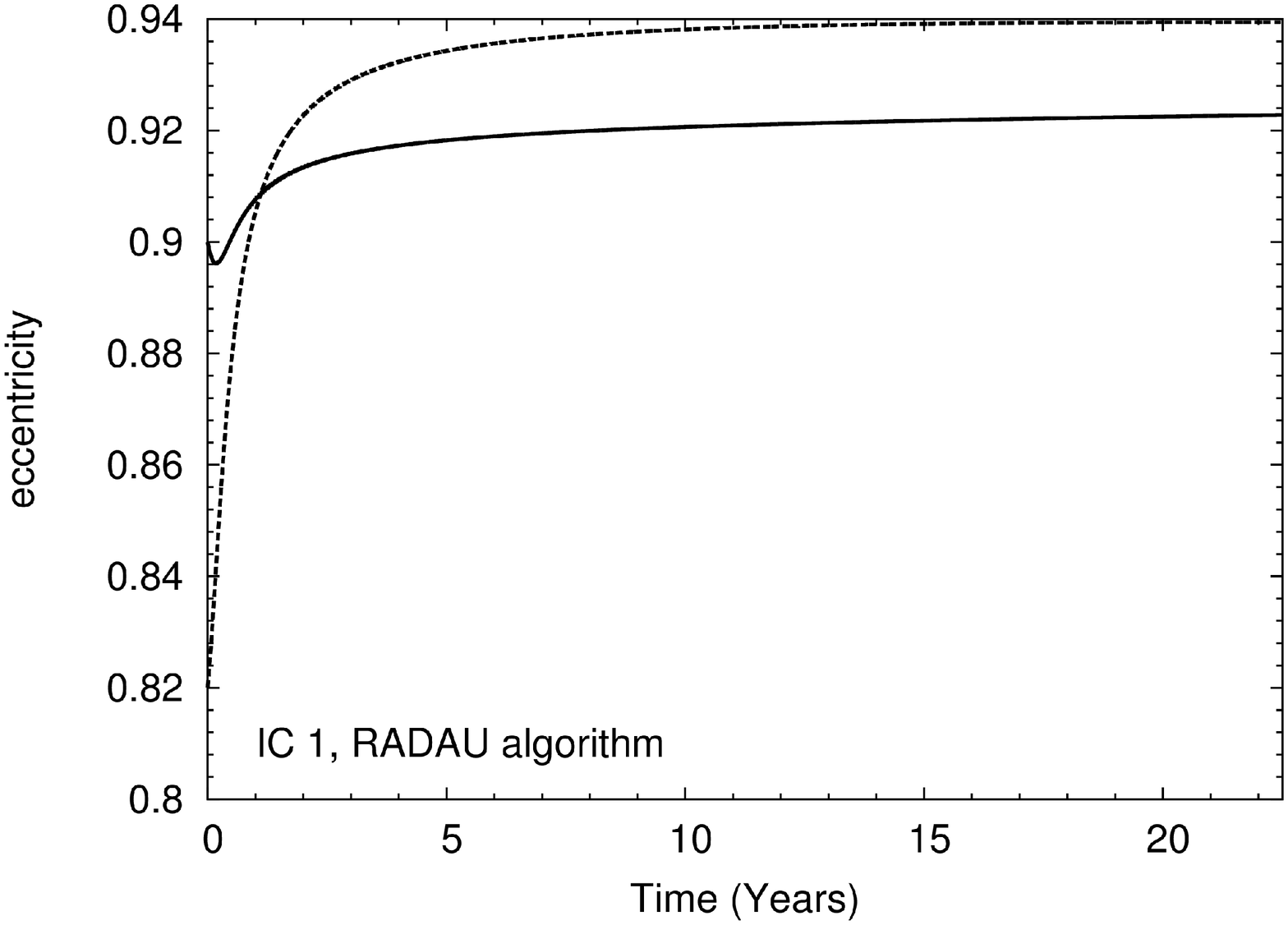}}}
}
\mbox
{
{\scalebox{0.185}{\includegraphics{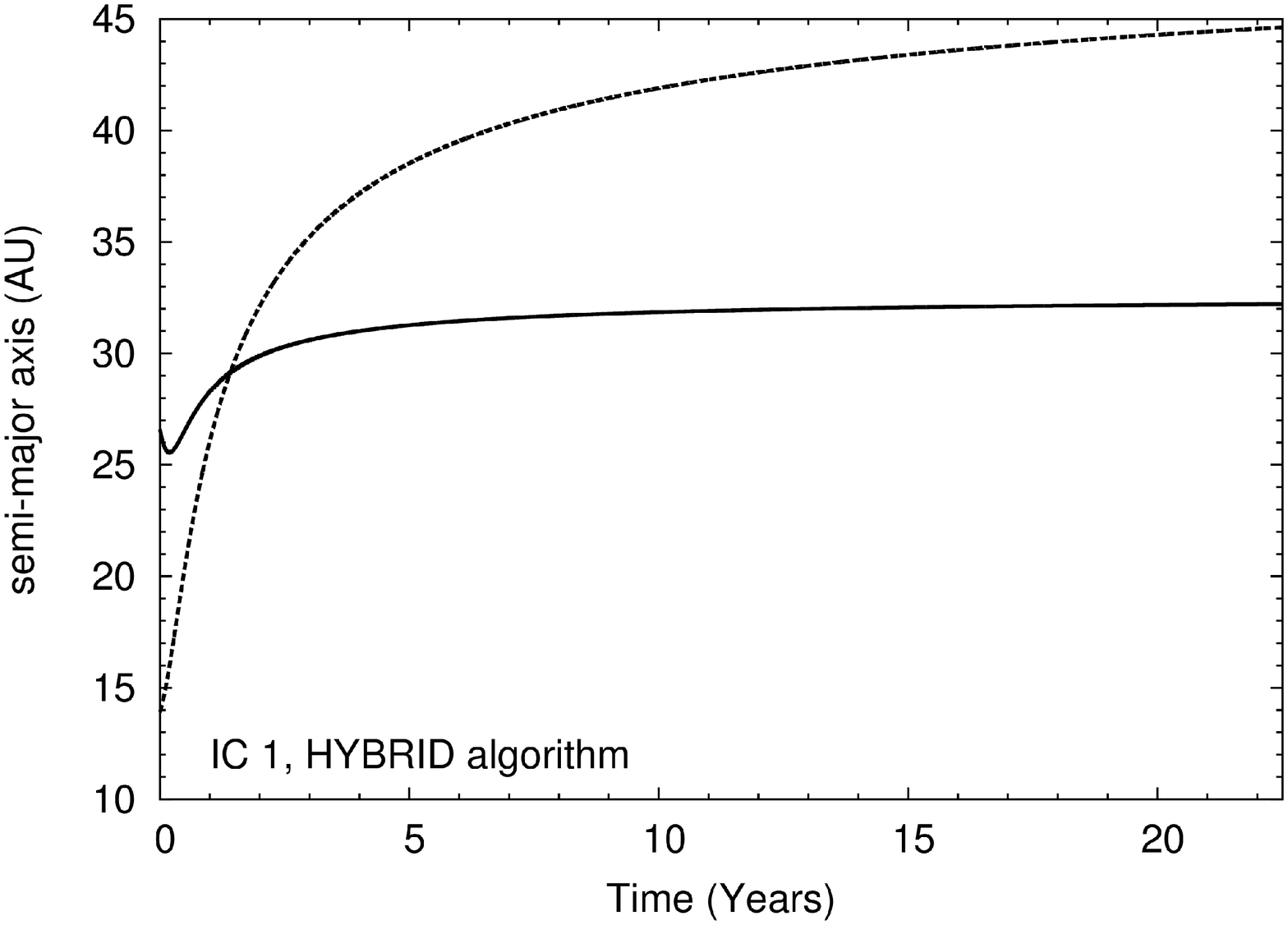}}}
{\scalebox{0.185}{\includegraphics{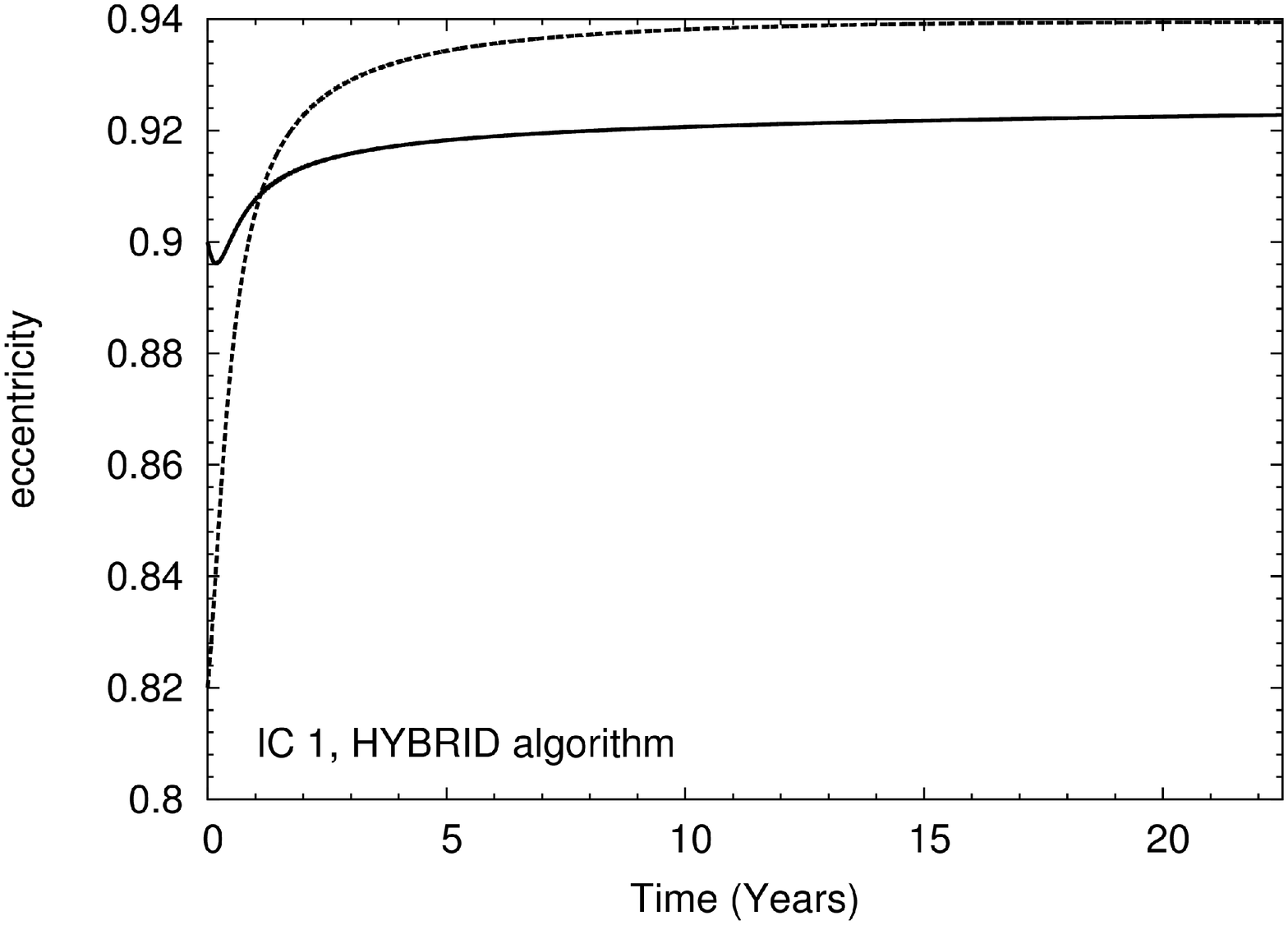}}}
}
\caption{Results from numerical tests considering the orbits of the two
companions as shown in Fig.~\ref{neworbits}. Here we consider a initial
condition (IC 1) where the outer companion has been assigned an initial
eccentricity of 0.90 (black dot in Fig.~\ref{neworbits}). All other parameters
are as shown in Table \ref{bestfitparamtable}. For the RADAU algorithm (top
panels) we used an initial time step of 0.001 days. For the HYBRID algorithm (bottom panels) we used a constant time step of 1 day. For both algorithms we used an accuracy parameter of $10^{-12}$. Orbital elements are sampled every 
day and are plotted until the ejection time of RZ Dra(AB)C at around $22$ years.}
\label{orbittest1}
\end{figure}

The results of our simulations can be seen in Figure~\ref{original}. It is immediately apparent that the system as proposed in Yang et al., 2010, is extremely unstable - with mean lifetimes ranging between $\simeq 100$ years 
and $\simeq 1600$ years. Indeed, the longest lived of the 126075 systems tested fell apart after just 173000 years. A remarkable $21\%$ of the test architectures were unstable on timescales of less than 100 years (25805 of the 126075 systems tested). Over 88\% of the systems fell apart within 1000 years (111225 of the systems tested). Whilst it is clear from the figure that the stability of the proposed system does increase somewhat as the separation of the companions is increased, this effect is clearly insufficient to allow the proposed companions to survive on sufficiently long timescales for their existence to be reasonable.
The likelihood of, by chance, observing the RZ Dra system within the last thousand years (or even the last hundred thousand years) before the destruction or ejection of the companions seems small. 

We have tested the accuracy of our numerical computations by comparing the results from several HYBRID integrations with results obtained from accurate RADAU integrations. In Fig.~\ref{orbittest1} we show the results of integrating an orbit (black dot in Fig.~\ref{neworbits}) of the outer companion with initially high eccentricty. For the chosen integration time step and accuracy parameters we conclude that the HYBRID integrations are reliable since the two orbits seem to follow the same time evolution until the inner companion is ejected after some 20 years. Orbits integrated with the RADAU algorithm are generally considered as producing reliable results. We carried out similar spot tests for other initial conditions with similar outcome.

As was the case for a number of 
other proposed circumbinary systems \citep[HU Aqr]{HUAqr, HUAqr2}; HW Vir \citep{HWVir}; and NSVS 14256825 \citep{NSVS}), the RZ Dra system proposed by Yang et al. (2010) does not stand up to dynamical scrutiny. If massive companions do exist within that system, they must clearly move on orbits far different to those proposed in the discovery work. The likelihood of observing the RZ Dra system in the last 1,000 years (or even 100,000 years) before the destruction or ejection of its companions, by chance, is vanishingly small 
given the host system's lifetime extends several hundred million years.

\begin{figure}
\includegraphics[scale=0.28]{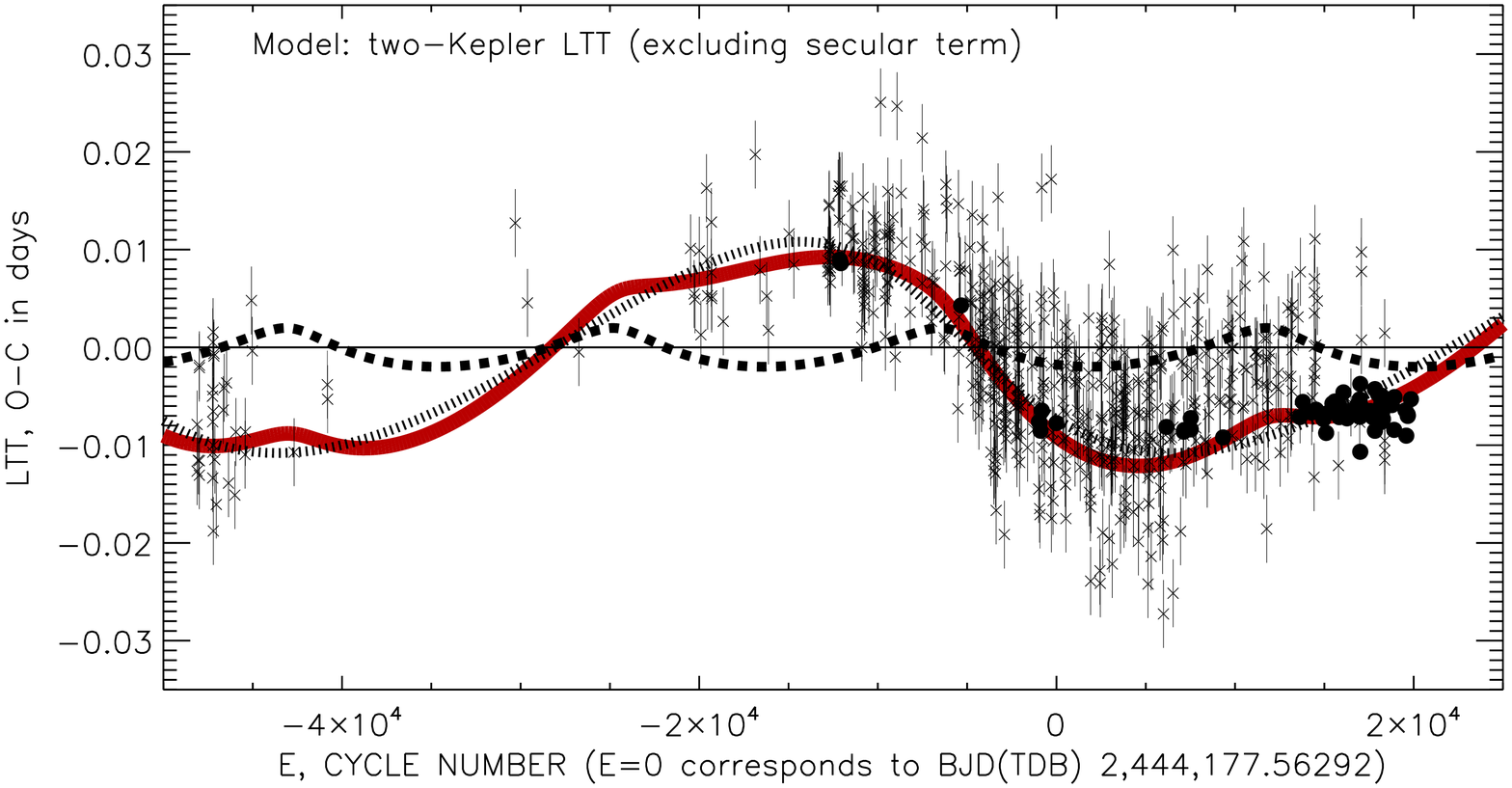}
\includegraphics[scale=0.28]{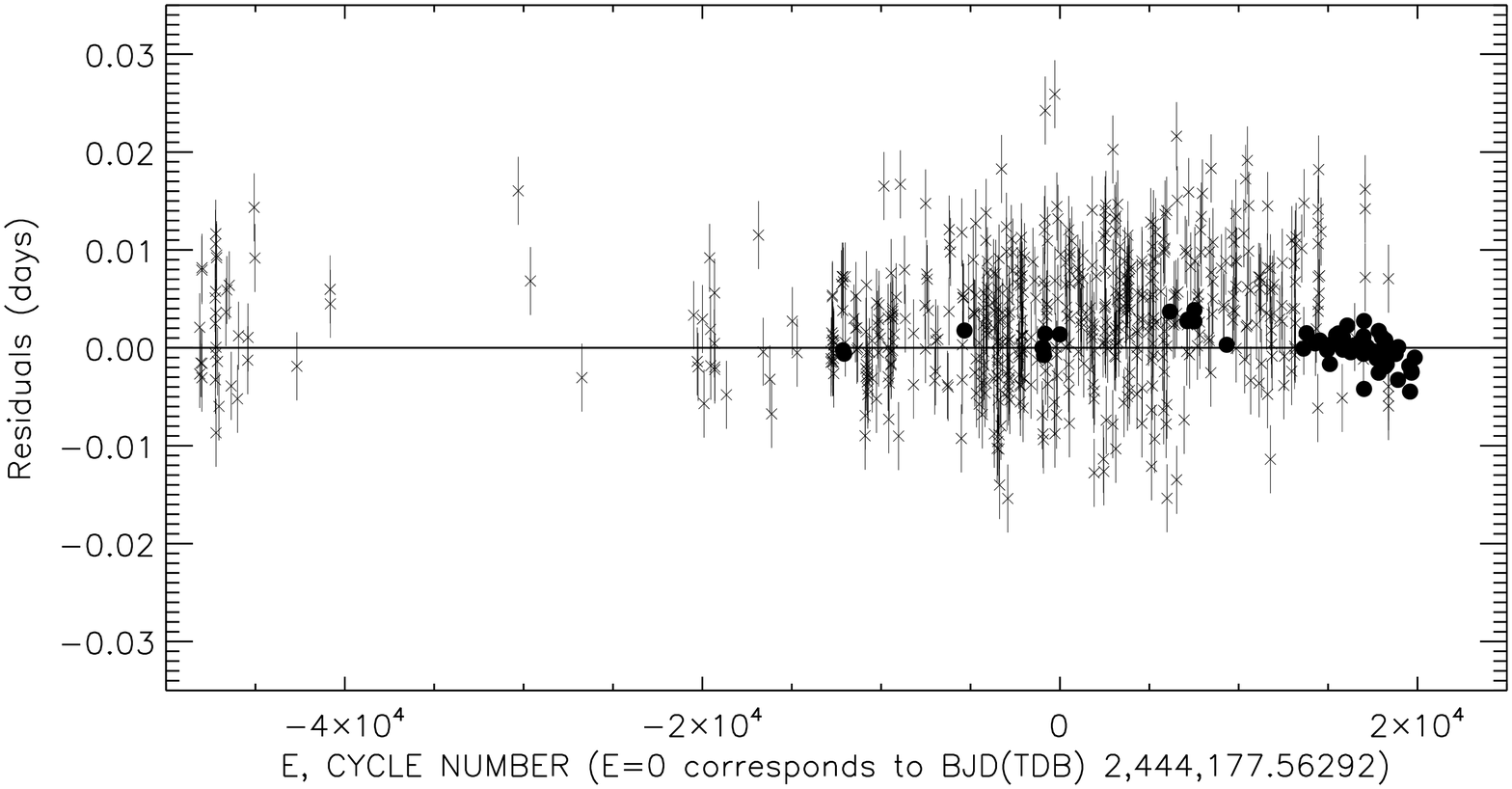}
\caption{Best-fit two-Kepler LTT model without secular parameter with reduced $\chi_{r}^2=12.7$. The best-fit model parameters are shown in Table \ref{bestfitparamtable}. In the top panel the LTT signal with smaller/larger amplitude corresponds to the inner/outer companion. The bottom panel shows the residuals between the data and best-fit model. We note that, judging by eye, a slight asymmetry is present in the residuals with $O-C > 0$ for $E>0$.}
\label{RZDRABestFitNoSec}
\end{figure}

\section{Data analysis and LTT model}
\begin{table*}
\centering
\begin{tabular}{lcccc}
\hline
Parameter && \multicolumn{2}{c}{two-LTT} & Unit \\ [1.5mm] \cline{3-4} \\ [-2.0ex]
                         && $\tau_{1}~(i=1)$                 & $\tau_{2}~(i=2)$           &   \\ 
\hline
$\chi_{r}^2$            &&  \multicolumn{2}{c}{12.75}                                     &   -   \\
\hline
RMS            &&  \multicolumn{2}{c}{535}                                                & seconds   \\
\hline
$T_0$                    &&  \multicolumn{2}{c}{$2,444,177.56292 \pm 1.1\times 10^{-5}$}  & BJD \\
$P_0$                    &&  \multicolumn{2}{c}{$0.5508761461 \pm 5.1 \times 10^{-10}$}   & days \\

$a_{b,i}\sin I_{b,i}$                && $0.34 \pm 0.03$     & $1.90 \pm 0.05$      & AU   \\

$e_{b,i}~~(\textnormal{or}~e_{1,2})$ && $0.82 \pm 0.27$     & $0.62 \pm 0.42$      & -    \\
$\omega_{b,i}$                       && $129 \pm 33$        & $325 \pm 56$         & deg.  \\
$T_{b,i}$                            && $2,443,429 \pm 212$ & $2,441,573 \pm 325$  & BJD \\
$P_{b,i}~~(\textnormal{or}~P_{1,2})$ && $11479 \pm 5883$    & $28807 \pm 8921$     & days \\
\hline
\hline
$m_{i}\sin I_{i}$                   && $0.07 \pm 0.01$              & $0.40 \pm 0.12$               & $M_{\odot}$ \\
$a_{i}\sin I_{i}$                    && $13.9 \pm 5.31$     & $26.6 \pm 5.81$      & AU   \\
$e_{i}$                              && $0.82 \pm 0.27$    & $0.62 \pm 0.42$       & -    \\
$\omega_{i}$                         && $309\pm 33$         & $145 \pm 56$         & deg  \\
$P_{i}$                              && $11479 \pm 5883$    & $28807 \pm 8921$     & days \\
\hline
\end{tabular}
\caption{Best-fit parameters for the LTT orbits of RZ~Dra corresponding 
to Fig.~\ref{RZDRABestFitNoSec}. Subscripts $1,2$ refer to the circumbinary 
companions with $i=1$, the inner, and $i=2$, the outer, companions.  RMS 
measures the root-mean-square scatter of the data around the best fit model. Uncertainties 
for the companion's semi-major axis ($a_{i}\sin I_{i}$) have been derived from 
Kepler's 3rd law using the largest uncertainty. The mass of the companions were held 
constant in the error propagation.}
\label{bestfitparamtable}
\end{table*}
We have used the same timing data\footnote{We note that new timing data exists as published by \cite{Erdem2011}. These data points were not included in the analysis presented here.} set as \cite{YangEtAl2010} along with their adopted timing precisions. In order to remove systematic offsets in the measured times of minimum light we have transformed the HJD (Heliocentric Julian Dates) time stamps (all assumed to be in the UTC time standard) to BJD (Barycentric Julian Dates) times in the TDB (Barycentric Dynamical Time) time standard using the transformation routines in \cite{Eastman2010}. Timing measurements earlier than JD~2,433,266.0 are limited by the DE405 JPL Planetary Ephemeris\footnote{http://ssd.jpl.nasa.gov/?horizons} \citep{Giorgini1996} and hence were transformed by omitting the light-travel time corrections from the Einstein and Shapiro effects (J. Eastman, priv. comm.). Since these effects contribute with timing precision to much less than one second, we judge them to be negligible in this work and thus have been omitted.

For an idealised, unperturbed and isolated binary system, the linear 
ephemeris of future/past (primary) mid-eclipse events can be 
computed from
\begin{equation}
T_{C}(E) = T_{0} + P_{0}E,
\label{ephemeriseq}
\end{equation} 
\noindent
where $E$ denotes the cycle number, $T_{0}$ is the reference epoch of some primary eclipse, and 
$P_{0}$ measures the eclipsing period of RZ~Dra. A linear regression on the 
680 recorded eclipse times allows the determination of $P_{0}$ to a high precision. In this work, we chose to place the reference epoch at the same date as in \cite{YangEtAl2010}. This is relatively close to the middle of the overall data set and would avoid parameter correlation between $T_{0}$ and $P_{0}$ during the fitting process. In the following we briefly outline the LTT model as used in this work.

\subsection{Analytic LTT model}

The model used in this work is similar to the formulation of a single 
light-travel time orbit introduced by \cite{Irwin1952}. In this model, 
the two binary components are assumed to represent one single object with a total mass equal to the sum of the masses of the two stars. If a circumbinary companion exists, then the combined binary mass follows an orbit around the system barycenter. The eclipses are then given by Eq.~\ref{ephemeriseq}. This defines the LTT orbit of the binary. The underlying reference system has its origin at the centre of the LTT orbit.

Following \cite{Irwin1952}, if the observed mid-eclipse times exhibit a 
sinusoidal variation (due to an unseen companion), then the 
quantity $O-C$ measures deviation from the linear ephemeris possibly attributable to the light-travel time effect and is given by
\begin{equation}
(O-C)(E) = T_{O}(E) - T_{C}(E) = \sum_{i=1}^2\tau_{i},
\label{omc}
\end{equation}
\noindent
where $T_{O}$ denotes the measured time of an observed mid-eclipse, 
summed over $i$ circumbinary companions. We here use the notion ``eclipse timing diagram'' when plotting the quantity ``$O-C$'' as a function of time. 

We note that $\tau_1+\tau_2$ is the combined LTT effect from two {\it separate} two-body LTT orbits. The quantity $\tau_i$ is given by the following expression for each companion 
\citep{Irwin1952}:
\begin{equation}
\tau_{i} = K_{b,i}\Big[\frac{1-e_{b,i}^2}{1+e_{b,i}\cos f_{b,i}} 
\sin (f_{b,i}+\omega_{b,i}) + e_{b,i}\sin \omega_{b,i}\Big ],
\label{taueq}
\end{equation}
\noindent
where $K_{b,i} = a_{b,i}\sin I_{b,i}/c$ is the semi-amplitude of the 
light-time effect (in the $O-C$ diagram) with $c$ measuring the speed of 
light and $I_{b,i}$ is the line-of-sight inclination of the LTT orbit 
relative to the sky plane, $e_{b,i}$ the orbital eccentricity, $f_{b,i}$ 
the true anomaly and $\omega_{b,i}$ the argument of pericenter. The 5 
model parameters for a single LTT orbit are given by 
the set $(a_{b,i}\sin I_{b,i}, e_{b,i}, \omega_{b,i}, T_{b,i}, 
P_{b,i})$. The time of pericentre passage $T_{b,i}$ and orbital period 
$P_{b,i}$ are introduced due to expressing the true longitude as a 
time-like variable via the mean anomaly $M = n_{b,i}(T_{O}-T_{b,i})$, 
with $n_{b,i} = 2\pi/P_{b,i}$ denoting the mean motion of the combined 
binary in its LTT orbit.  Computing the true anomaly as a function of 
time (or cycle number) requires the solution of Kepler's equation. We 
refer to \cite{HinseEtAl2012a} for more details.

\section{Method of least-squares fitting and parameter uncertainties}
\begin{figure*}
\includegraphics[scale=0.75]{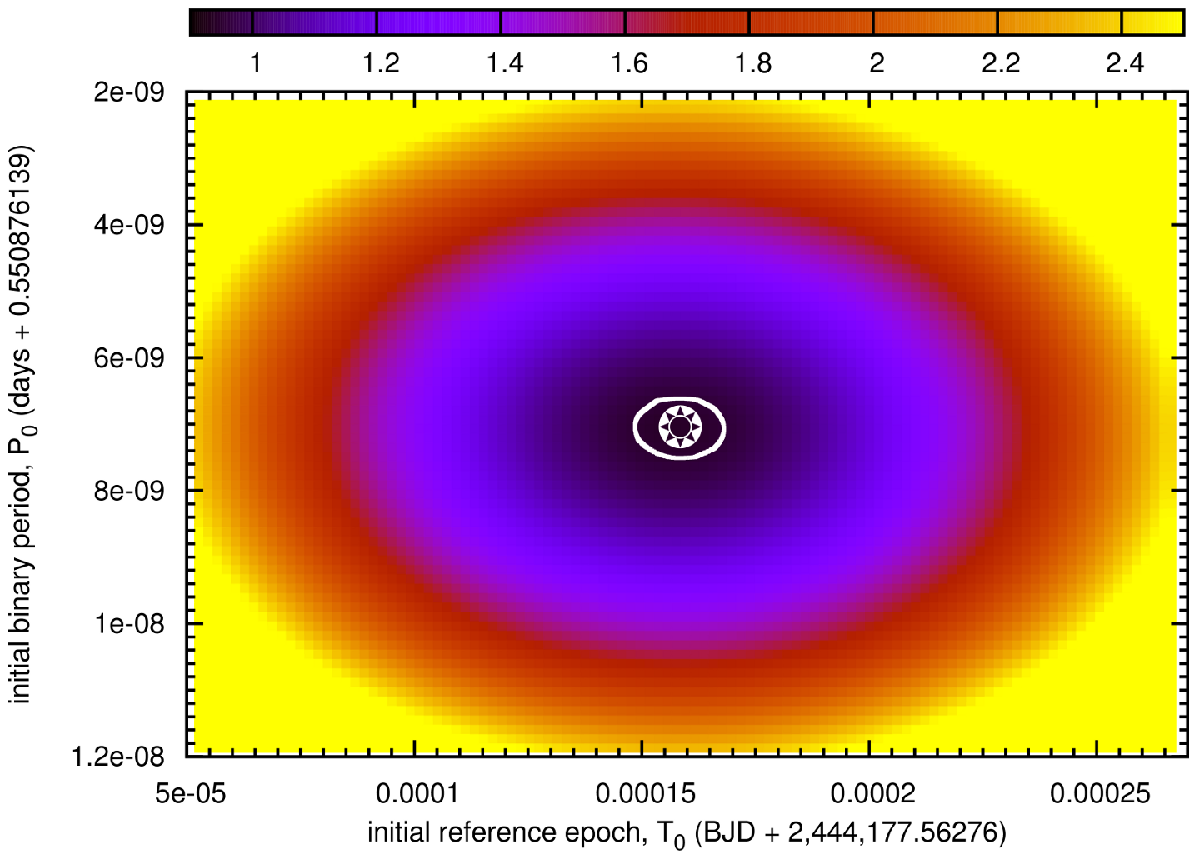}
\includegraphics[scale=0.75]{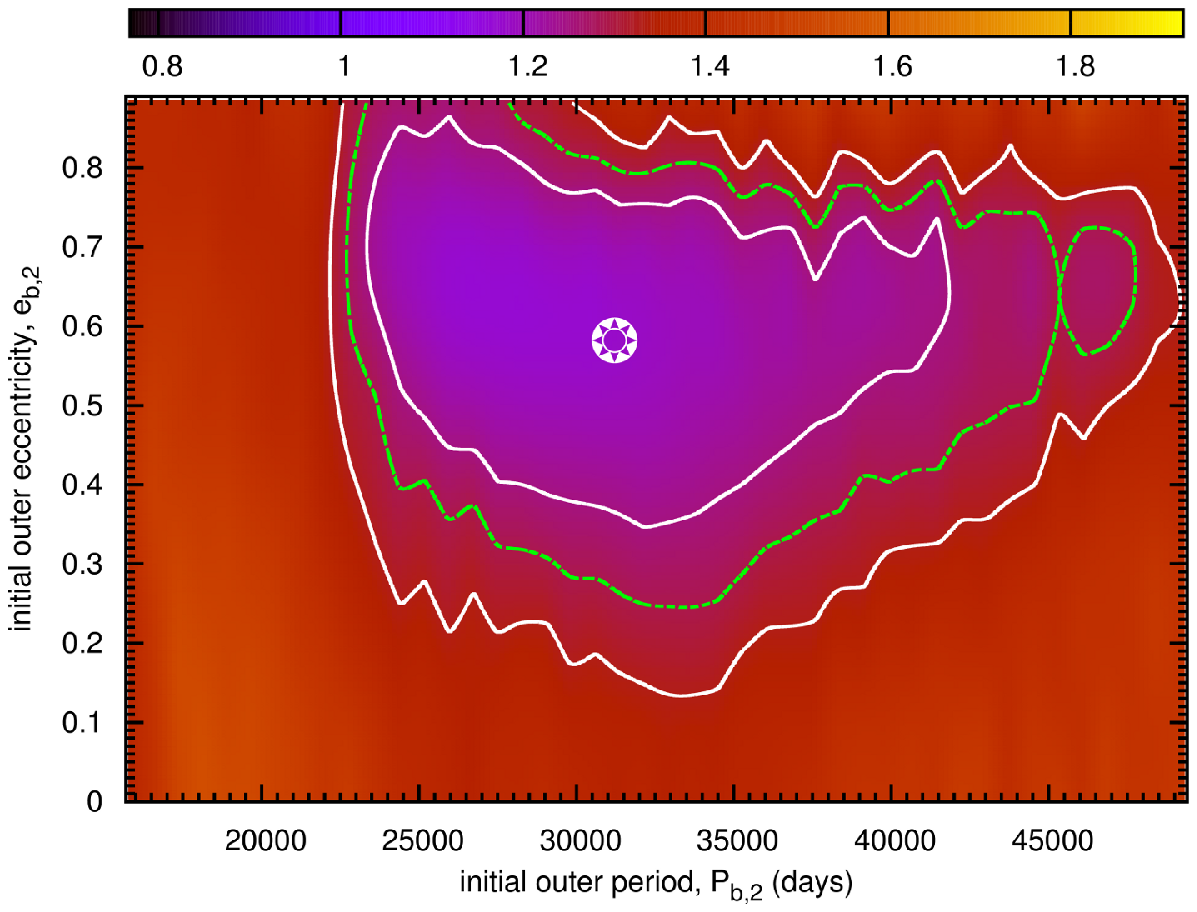}
\vspace{0.0cm}
\caption{Two-dimensional parameter scans of the $\chi^2_{r}$ statistic in a neighborhood around the best-fit model (marked with an asterisk symbol) considering the two-companion model. The colorbar is scaled with the value of our best-fit model representing 1. Countour lines show the 1-$\sigma$ (left panel) and 1-,2-,3-$\sigma$ (right panel) confidence level curves around the best-fit model. {\it See electronic version for colors}.}
\label{scan1}
\end{figure*}

The overall procedure of finding a best-fit model follows the methodology as outlined in \cite{HinseEtAl2012a} using a least-squares minimisation algorithm as implemented in \texttt{MPFIT} \citep{Markwardt2009}. We considered several models. First we started to fit a single-companion LTT model to the data with 7 freely varying parameters (680-7 degrees of freedom). Considering a few ten thousand initial random guesses (all parameters randomised as in \cite{HinseEtAl2012a}) we found the reduced best-fit statistic $\chi^2_{r}$ to be no less than $10^4$ indicating a one-companion model to be inadequate to describe the data. We then fitted for a two-companion model with all (12) parameters allowed to vary freely (680-12 degrees of freedom). A significantly better fit was found with $\chi^2_{r}\simeq 13$. To investigate the possibility of a secular trend in the timing data set (representing a linear period decrease by mass transfer) we conducted an independent best-fit model search by adding an additional term to Eq.~\ref{omc} which is quadratic in time (or cycle number). However, upon inspecting the residual plots we were not convinced about the necessity of adding an additional free parameter. We judged that no obvious secular parabolic trend was present in the dataset. For a one-companion model with a quadratic trend we refer to \cite{Erdem2011}. Our best-fit two-companion model and resulting best-fit parameters are shown in Fig.~\ref{RZDRABestFitNoSec} and Table \ref{bestfitparamtable}.

During this study (in conjunction with previous studies, e.g. \cite{HWVir}) we found that the formal parameter errors as obtained from the best-fit co-variance matrix are too optimistic. We have therefore determined parameter errors from Monte Carlo Bootstrap experiments \citep{GozdziewskiEtAl2012}. For each bootstrap dataset we used the best-fit model as our initial guess and recorded the resulting parameters. We generated $10^6$ bootstrap simulations. From the resulting distributions we determined the 1-$\sigma$ uncertainty in each parameter to be the 68.2\% percentile interval. A graphical representation of our confidence intervals is given in Fig.~\ref{scan1} showing a two-dimensional joint-confidence parameter scan of the $\chi^2_{r}$ statistic. Here we consider the dependency between the period and eccentricity parameters of the outer companion as well as the reference epoch and the binary period. The 1-$\sigma$ confidence level agrees well with the outcome of our bootstrap experiment.

\section{Dynamical stability of our new best-fit model}

Having derived a new Keplerian model for the observed timing variations in the RZ~Dra system, we once again carried out a detailed dynamical study of the proposed orbits. Following the same procedure as described in section 2, we held the initial orbital elements of RZ~Dra (AB)~C fixed at their nominal best-fit values, and created 126075 test systems in which the initial semi-major axis, eccentricity, mean anomaly and argument of periastron for RZ~Dra (AB)~D were systematically varied across the $\pm 3\sigma$ range around their nominal best-fit values. Once again, we considered 41 discrete values of $a$ and $e$, 15 unique values of $\omega$ and 5 unique values of $M$. Due to the large values of $a$ and $e$ allowed by the $\pm 3\sigma$ range, the initial apastron distance for RZ Dra (AB) D can range as high as $88.06$ AU - and so we set the ejection distance as $150$ AU in this case, so that objects are only considered to be ejected if they have experienced significant orbital perturbations.

\begin{figure*}
\sidecaption
\includegraphics[width=120mm]{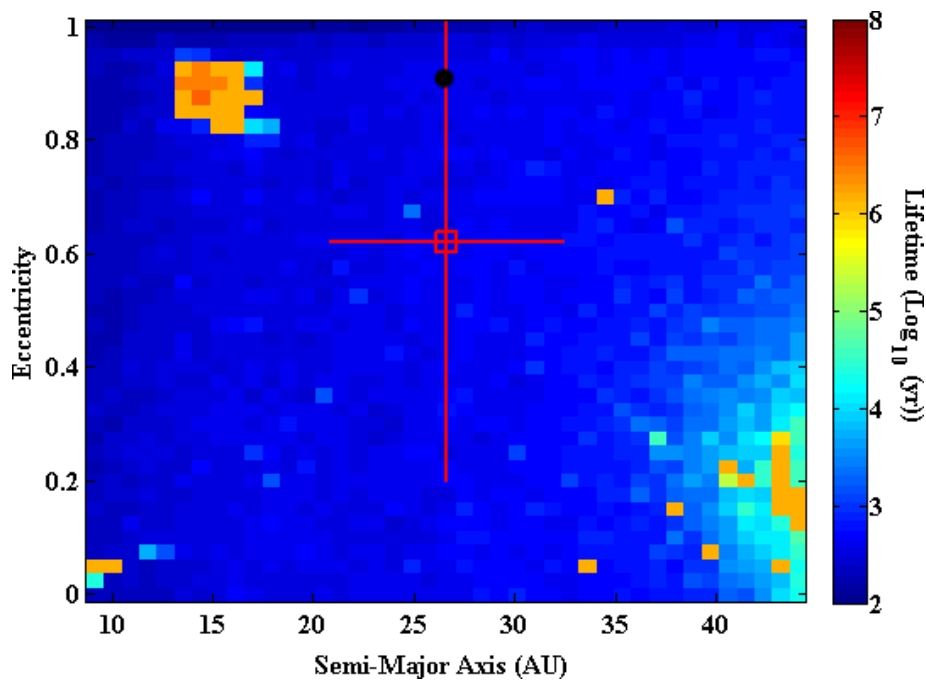}
\caption{Dynamical stability of the RZ Dra system, as a function of the semi-major axis, $a$, and eccentricity, $e$ of RZ Dra (AB) D, for the alternative solution presented in this work. The nominal best-fit orbit for RZ Dra (AB) D is located in the centre of the red box, at $a = 26.6$ AU and $e = 0.62$. The $1\sigma$ uncertainties on the semi-major axis and eccentricity are shown by a crosshair. Almost all the tested models, particularly those within the $\pm 1 \sigma$ uncertainties, are dynamically unstable on remarkably short timescales. A few regions display stability on timescales of around a million years, but these are all located at the extreme edges of the plot, almost $\pm 3 \sigma$ away from the nominal best fit orbit. The black dot is the location of a test orbit as discussed in section 2. {\it See electronic version for colors}.}
\label{neworbits}
\end{figure*}

The results of our simulations are shown in Figure~\ref{neworbits}. As was the case for the companions proposed by Yang et al., the new model parameters leads to extremely unstable orbits. The entire region spanned by the $\pm 1\sigma$ uncertainties features mean lifetimes measured in thousands of years, and the only (very small) regions where the stability is measured in hundreds of thousands of years, or even just over a million years, are located a large distance from the nominal best-fit model. 

On the basis of these results, it seems reasonable to conclude that the observed variations in the timing of the eclipses in the RZ~Dra system are not the result of the gravitational influence of two unseen companions - such companions would be so massive, and move on such extreme orbits, that they would destabilise one another on timescales of hundreds of years - far shorter than the typical lifetime of the host binary system. {\bf We would like to remind the reader that the nature of near-contact binaries is not well-known.  Currently there is a common consensus that near-contact binaries are intermediate objects observed between two contact phases. In contact binaries mass and energy exchange between the two components can be significant  which may and should affect the interior structure of the stars. The orbital distance between the two stellar surfaces is just a few percent of the stellar radii. We therefore anticipate that stellar mass-transfer is possible for via a flare that moves matter from one component to the other. This could cause a sudden period change of the eclipsing binary resulting in the observation of ETVs.}

\section{Discussion \& Conclusion}

Based on measurements of primary and secondary eclipse times \cite{YangEtAl2010} proposed to interpret the timing variation due to two circumbinary companions. The two bodies they propose are very low mass stars with minimum mass $0.07~M_{\odot}$ for the inner and $0.18~M_{\odot}$ for the outer companion. From their best-fit model the inner companion's apocenter distance is at $\simeq 18$ AU and the outer companion's pericenter distance is $\simeq 17$ AU, implying a crossing-orbit architecture. We tested the dynamical feasibility of the proposed best-fit orbits by exploring a large grid of initial conditions using direct $n$-body integrations. The results of our simulations exploring the $(a,e)$-space are shown in Fig.~\ref{original}. All of the tested orbits were highly unstable, and resulted in either the break-up of the system or collisions between the two bodies. We found that the mean lifetimes range between 100 and 1600 years. {\bf In this work we have not considered mutually inclined orbits between the two companions. For the interested reader, we would like to point to the works of \citep{HUAqr,HinseEtAl2012b,QSVir,BD+20,Wittenmyer2014} who investigated the effect of mutually inclined orbits leading to the result of little improvement to the overall dynamical stability. However, the only exception to this finding were those scenarios for which the two planets were placed on anti-coplanar orbits - in other words, where they moved in the same plane, but with a mutual orbital inclination of 180 degrees. This setup predictably led to extremely stable systems whenever the two planets were not placed on orbits that crossed one another (which led to extreme instability). We judge that such an orbital architecture, whilst of theoretical interest, seems highly physically implausible, and so need not be considered further at this time.}

We then searched for and determined a new best-fit model using the complete set of timing measurements transformed from HJD to BJD time standard. We found a new best-fit model with parameters shown in Table \ref{bestfitparamtable}. Compared to the orbital parameters given in \cite{YangEtAl2010} our new model assigned higher eccentricities to the orbits of the companions and increased the mass of the outer companion to $0.4~M_{\odot}$. We then tested the our new model parameters for dynamical stability and found life-times of only a few hundred years (Fig.~\ref{neworbits}).

On the basis of our dynamical simulations we conclude the two companion hypothesis around RZ Dra does not stand up to scrutiny. At this point we would like to emphasize on the robustness of our stability analysis. First we have carried out tests which demonstrate that the results obtained from HYBRID integrations are reliable. Second, our dynamical setup replaces the two binary components as a single massive object. In that sense our setup considers a system which favours dynamical stability by ignoring the gravitational perturbations from a extra (significant in mass) body. If no stable orbits are found in this simplified system it is generally hard to conceive how stable orbits are ensured by adding an additional pertubing force to the system. From an intuitive point of view adding more perturbing forces will increase the effect and possibility of chaos and hence favours orbital instabilities. Another aspect of replacing the two binary components by a single body is an issue of being consistent with the LTT model which assumes the binary to be a single massinve object.

It is clearly important, therefore, to consider what other mechanisms could account for the observed variations. In Fig.~\ref{RZDRABestFitNoSec} we have some indication by an asymmetric distribution of the residuals of additional effects that might be causing a period change. The orbital periods of many binary systems have varied due to some combination of secular and/or cyclical variations. Generally, the quasi-sinusoidal timing variations could be produced by at least three physical causes: (1) apsidal motion in an elliptical orbit of the binary, (2) a light-travel-time (LTT) effect (or several) due to additional companion(s), or (3) cyclical changes of magnetic activity of the component stars with deep convective envelope. The secular variations can be interpreted as being caused by either mass transfer between the two component stars or by angular momentum loss (AML).

In case of RZ Dra, because the binary has a circular orbit, the cyclical variations cannot be described by apsidal motion. Further, \cite{YangEtAl2010} ruled out the magnetic activity cycles because the variations of the gravitational quadrupole moment ($\Delta Q$) are two orders of magnitude smaller than typical values of $10^{51} - 10^{52}$ for close binaries. This finding is further supported by a recent study \citep{Lanza2006} which indicates that the magnetic mechanism (Applegate model) is not sufficiently effective to explain the period modulation of close binaries with a late-type secondary.

However, the eclipsing pair of RZ Dra is a semi-detached binary with the less massive secondary filling its inner Roche lobe \cite{YangEtAl2010}. In such semi-detached binaries, a secular variation (quadratic term) could be produced through a mass transfer from the secondary to the primary star, AML due to a magnetic stellar wind, or the combination of the two aforementioned mechanisms. The mass transfer cause a period increase (upward parabola), while the AML a period decrease (downward parabola). In this work we considered a quadratic + two-LTT model (see Section 4), but we were not able to convincingly detect either one of these trends in the residuals. However, a secular variation may be hidden in the timing data set and this system may be in a weak phase of mass transfer. We refer to \cite{Erdem2011} who considered a quadratic + 1-LTT model.

Furthermore, the presence of systematic residuals shown in Fig.~\ref{RZDRABestFitNoSec} might be due to an apparent phase shift of the 
real conjunctions due to asymmetrical eclipse minima originating from starspot activity \citep{LeeEtAl2009b}. The effects of starspots on timing measurments of eclipsing binaries were also studied by \cite{WatsonDhillon2004}.

At present we can not rule out that most of the timing measurements have been underestimated and hence the plotted errorbars in Fig.~\ref{RZDRABestFitNoSec} could be much larger than stated in the literature. To obtain an idea of the timing uncertainty we scrutinized table 4 in \cite{YangEtAl2010} and noticed 
the visual recording of the same secondary eclipse (presumably by two observers) on the night of HJD~2,442,984 (July 24, 1976; we refer to the VO ASCII data file available on-line). The first observer measured the secondary eclipse to be at HJD~2,442,984.637 and the second observer measured the same eclipse event to occur at HJD~2,442,984.641. These observations suggests a larger uncertainty since the time of minimum light from the visual observations differ by more than 5 minutes (over 300 seconds). This assumes, of course, that the entry in the corresponding VO file is neither a duplicated entry or typing error. If times of minimum light truely have a precision of 0.003 days, then most of the variation seen in Fig.~\ref{RZDRABestFitNoSec} would be within the noise-level of timing estimates. In general, \cite{Eastman2010} recommend that uncertainties of at least 60 seconds should be used for the timing precision, if the time standard has not been specified explicitly.

We therefore encourage future follow-up observations of eclipsing binaries to obtain as precise timing measurements as possible for RZ Dra and other systems mainly characterised via visual measurements of the minima. Several monitoring programs are currently running or in planning \citep{Sybilski2010, PribullaEtAl2012} within the context of searching for circumbinary companions of planetary, sub-stellar and stellar nature.

\begin{acknowledgements}
We would like to thank the anonymous referee for improving this research paper. Research by TCH is carried out at the Korea Astronomy and Space Science 
Institute (KASI) under the KRCF (Korea Research Council of Fundamental Science 
and Technology) Young Scientist Research Fellowship Program. Numerical
simulations were carried out on ``Pluto'' Computing Cluster at KASI and the SFI/HEA Irish Centre for High-End Computing (ICHEC). TCH, JWL, CUL \& JHP acknowledges support from KASI registered under grant number 2013-9-400-00/2014-1-400-06. Astronomical research at Armagh Observatory is funded by the Department of Culture, Arts and Leisure (DCAL). JPM is supported by Spanish grant AYA 2011-26202. This work was supported by iVEC through the use of advanced computing resources located at the Murdoch University, in Western Australia. This research has made use of NASA's Astrophysics Data System (ADS).
\end{acknowledgements}


\begin{thebibliography}{}



\bibitem[Almeida et~al.(2011)]{Almeida2011} 
Almeida, L.~A., Jablonski, F., 2011, IAUS, 276, 495

\bibitem[Almeida et~al.(2013)]{Almeida2013} 
Almeida, L.~A., Jablonski, F., Rodrigues, C.~V., 2013, \apj, 766, 11

\bibitem[Beuermann et~al.(2010)]{BeuermannEtAl2010} 
Beuermann, K., Hessman, F.~V., Dreizler, S. et~al., 2010, A\&A, 521, 60

\bibitem[Beuermann et~al.(2011)]{BeuermannEtAl2011} 
Beuermann, K., Buhlmann, J., Diese, J. et al., 2011, A\&A, 526, 53

\bibitem[Beuermann et~al.(2013)]{BeuermannEtAl2013} 
Beuermann, K., Dreizler, S., Hessman, F.~V., 2013, A\&A, 555, 133

\bibitem[Borkovits \& Heged{\"u}s (1996)]{BoHe1996} 
Borkovits, T., \& Heged{\"u}s, T., 1996, A\&AS, 120, 63

\bibitem[Chambers(1999)]{Mercury} 
Chambers, J.~E., 1999, \mnras, 304, 793 

\bibitem[Chauvin et~al.(2007)]{Chauvin2007} 
Chauvin, G., Lagrange, A.-M., Udry, S., Mayor, M., 2007, A\&A, 475, 723

\bibitem[Deeg et~al.(2000)]{DeegEtAl2000} 
Deeg, H.~J., Doyle, L.~R., Kozhevnikov, V.~P. et~al., 2000, A\&A, 358, 5

\bibitem[Doyle \& Deeg(2004)]{DoyleDeeg2004} 
Doyle, L.~R., Deeg, H.-J., 2004, IAUS, 213, 80

\bibitem[Doyle et~al.(2011)]{DoyleEtAl2011} 
Doyle, L.~R., Carter, J.~A., Fabrycky, D.~C. et~al., 2011, Science, 333, 1602

\bibitem[Duchene et~al.(2007)]{DucheneEtAl2007} 
Duch{\^e}ne, G., Bontemps, S., Bouvier, J., et~al., 2007, A\&A, 476, 229

\bibitem[Eastman et al.(2010)]{Eastman2010}
Eastman, J., Siverd, R., Gaudi, B.~S., 2010, PASP, 122, 935 

\bibitem[Erdem et~al.(2011)]{Erdem2011} 
Erdem, A., Zola, S., Winiarski, M., 2011, New Astronomy, 16, 6

\bibitem[Evans(1968)]{Evans1968} 
Evans, D.~S., 1968, QJRAS, 9, 388

\bibitem[Funk et~al.(2011)]{FunkEtAl2011} 
Funk, B., Eggl, S., Gyergyovits, M. et~al., 2011, EPSC-DPS, 1725

\bibitem[Funk et~al.(2012)]{FunkEtAl2012} 
Funk, B., Eggl, S., Gyergyovits, M., et~al., 2012, IAUS, 282, 446

\bibitem[Furlan et~al.(2007)]{FurlanEtAl2007} 
Furlan, E., Sargent, B., Calvet, N. et~al., 2007, \apj, 664, 1176

\bibitem[Giorgini et~al.(1996)]{Giorgini1996}
Giorgini, J.~D., Yeomans, D.~K., Chamberlin, A.~B. et~al., 1996, BAAS, 28, 1158

\bibitem[Go{\'z}dziewski et~al.(2012)]{GozdziewskiEtAl2012} 
Go{\'z}dziewski, K., Nasiroglu, I., S{\l}owikowska, A. et~al., 2012, \mnras, 425, 930

\bibitem[Hessman et~al.(2010)]{HessmanEtAl2010} 
Hessman, F.~V., Dhillon, V.~S., Winget, D.~E. et~al., 2010, astro-ph, arXiv:1012.0707H

\bibitem[Hinse et~al.(2012a)]{HinseEtAl2012a} 
Hinse, T.~C., Lee, J.~W., Haghighipour, N. et~al. 2012a, \mnras, 420, 3609

\bibitem[Hinse et~al.(2012b)]{HinseEtAl2012b}
Hinse, T.~C., Go{\'z}dziewski, K., Lee, J.~W. et~al., 2012b, \aj, 144, 34

\bibitem[Hinse et~al.(2014)]{HinseEtAl2014}
Hinse, T.~C., Lee, J.~W., Go{\'z}dziewski, K. et~al., 2014, \mnras, 438, 307

\bibitem[Horner et al.(2011)]{HUAqr} 
Horner, J., Marshall, J.~P., Wittenmyer, R.~A., \& Tinney, C.~G.\ 2011, 
\mnras, 416, L11 

\bibitem[Horner et~al.(2012a)]{NNSer}
Horner, J., Wittenmyer, R.~A., Hinse, T.~C., Tinney, C.~G., 2012a, \mnras, 425, 749

\bibitem[Horner et~al.(2012b)]{HWVir}
Horner, J., Hinse, T.~C., Wittenmyer, R.~A., Marshall, J.~P., Tinney, C.~G., 2012b, \mnras, 427, 2812

\bibitem[Horner et~al.(2013)]{QSVir}
Horner, J., Wittenmyer, R.~A., Hinse, T.~C. et~al., 2013, \mnras, 435, 2033

\bibitem[Horner et~al.(2014)]{BD+20}
Horner, J., Wittenmyer, R.~A., Hinse, T.~C., Marshall, J.~P., 2014, \mnras, 
439, 1176

\bibitem[Irwin(1952)]{Irwin1952} 
Irwin, J.~B., 1952, \apj, 116, 211

\bibitem[Irwin(1959)]{Irwin1959} 
Irwin, J.~B., 1959, \aj, 64, 149

\bibitem[Kostov et~al.(2013)]{KostovEtAl2013} 
Kostov, V.~B., McCullough, P., Hinse, T.~C. et~al., 2013, \apj, 770, 52

\bibitem[Lanza(2006)]{Lanza2006}
Lanza, A.~F., 2006, \mnras, 369, 1773

\bibitem[Lee et~al.(2009a)]{LeeEtAl2009a} 
Lee, J.~W., Kim, S.-L., Kim, C.-H., et~al., 2009a, \aj, 137, 3181

\bibitem[Lee et~al.(2009b)]{LeeEtAl2009b} 
Lee, J.~W., Youn, J.-H., Lee, C.-U. et~al., 2009b, \aj, 138, 478

\bibitem[Lee et~al.(2011)]{LeeEtAl2011} 
Lee, J.~W., Kim, S.-L., Lee, C.-U. et~al., 2011, \aj, 142, 12

\bibitem[Lee et~al.(2012)]{LeeEtAl2012} 
Lee, J.~W., Lee, C.-U., Kim, S.-L. et~al., 2012, \aj, 143, 34

\bibitem[Lee et~al.(2013a)]{LeeEtAl2013a} 
Lee, J.~W., Kim, S.-L., Lee, C.-U. et~al., 2013a, \apj, 763, 74

\bibitem[Lee et~al.(2013b)]{LeeEtAl2013b} 
Lee, J.~W., Hinse, T.~C., Park, J.-H., 2013b, \aj, 145, 100

\bibitem[Lim \& Takakuwa(2006)]{LimTakakuwa2006} 
Lim, J., Takakuwa, S., 2006, \apj, 653, 425

\bibitem[Markwardt(2009)]{Markwardt2009} 
Markwardt, C.~B., 2009, ASPCS, "Non-linear Least-squares Fitting in 
IDL with MPFIT", Astronomical Data Analysis Software and Systems XVIII,
eds. Bohlender, D.~A., Durand, D., Dowler, P.

\bibitem[Marshall et al.(2010)]{HR8799} 
Marshall, J., Horner, J., Carter, A., 2010, International Journal of Astrobiology, 9, 259 

\bibitem[Marzari et~al.(2009)]{MarzariEtAl2009} 
Marzari, F., Scholl, H., Th{\'e}bault, P., Baruteau, C., 2009, A\&A, 508, 1493

\bibitem[Mustill et~al.(2013)]{Mustill2013} 
Mustill, A., Marshall, J.~P., Villaver, E. et~al., 2013, \mnras, 437, 1404

\bibitem[Neuh{\"a}user et~al.(2007)]{Neuhauser2007} 
Neuh{\"a}user, R., Mugrauer, M., Fukagawa, M., Torres, G., Schmidt, T., 2007, A\&A, 462, 777

\bibitem[Ofir et~al.(2009)]{Ofir2009} 
Ofir, A., Deeg, H.~J., Lacy, C.~H.~S., 2009, A\&A, 506, 445

\bibitem[Orosz et~al.(2012a)]{OroszEtAl2012a}
Orosz, J.~A., Welsh, W.~F., Carter, J.~A. et~al., 2012a, Science, 337, 1511

\bibitem[Orosz et~al.(2012b)]{OroszEtAl2012b}
Orosz, J.~A., Welsh, W.~F., Carter, J.~A. et~al., 2012b, \apj, 758, 87

\bibitem[Parsons et~al.(2010)]{ParsonsEtAl2010} 
Parsons, S.~G., Marsh, T.~R., Copperwheat, C.~M. et~al., 2010, \mnras, 407, 2362

\bibitem[Potter et~al.(2011)]{PotterEtAl2011} 
Potter, S.~B. Romeo-Colmenero, E., Ramsay, G. et~al., 2011, \mnras, 416, 2202

\bibitem[Pribulla et~al.(2012)]{PribullaEtAl2012} 
Pribulla , T., Va{\v n}ko, M., Ammler-von Eiff, M. et~al., 2012, AN, 333, 754

\bibitem[Qian et~al.(2009)]{QianEtAl2009} 
Qian, S.-B., Dai, Z.-B., Liao, W.-P., et~al., 2009, \apj, 706, L96

\bibitem[Qian et~al.(2010a)]{QianEtAl2010a} 
Qian, S.-B., Liao, W.-P., Zhu, L.-Y., Dai, Z.-B., 2010a, \apj, 708, L66

\bibitem[Qian et~al.(2010b)]{QianEtAl2010b} 
Qian, S.-B., Liao, W.-P., Zhu, L.-Y. et~al., 2010b, \mnras, 401, L34

\bibitem[Qian et~al.(2011)]{QianEtAl2011} 
Qian, S.-B., Liu, L., Liao, W.-P. et~al., 2011, \mnras, 414, L16

\bibitem[Qian et~al.(2012a)]{Qian2012a} 
Qian, S.-B., Liu, L., Zhu, L.-Y. et~al., 2012a, \mnras, 422, 24.

\bibitem[Qian et~al.(2012b)]{Qian2012b}
Qian, S.-B., Zhu, L.-Y., Dai, Z.-B. et~al., 2012b, \apj, 745, 23

\bibitem[Reipurth(2000)]{Reipurth2000} 
Reipurth, B., 2000, \aj, 120, 3177

\bibitem[Robertson et al.(2012a)]{HD155358}
Robertson, P., Endl, M., Cochran, W.~D., et al., 2012a, \apj, 749, 39 

\bibitem[Robertson et al.(2012b)]{HD204313} 
Robertson, P., Horner, J., Wittenmyer, R.~A., et al., 2012b, \apj, 754, 50

\bibitem[Schwamb et~al.(2013)]{SchwambEtAl2013}
Schwamb, M.~E., Orosz, J.~A., Carter, J.~A. et~al., 2013, \apj, 768, 127

\bibitem[Schwarz et~al.(2011)]{Schwarz2011}
Schwarz, R., Haghighipour, N., Eggl, S., Pilat-Lohinger, E., Funk, B.,
2011, \mnras, 414, 2763

\bibitem[Sybilski et~al.(2010)]{Sybilski2010} 
Sybilski, P., Konacki, M., Koz{\l}owski, S., 2010, \mnras, 405, 657

\bibitem[van den Berk et~al.(2007)]{vandenBerkEtAl2007} 
van den Berk, J., Portegies Zwart, S.~F., McMillan, S.~L.~W., 2007, \mnras, 
379, 111

\bibitem[Watson \& Dhillon(2004)]{WatsonDhillon2004} 
Watson, C.~A. \& Dhillon, V.~S., 2004, \mnras, 351, 110

\bibitem[Welsh et~al.(2012)]{WelshEtAl2012}
Welsh, W.~F., Orosz, J.~A., Carter, J.~A. et~al., 2012, Nature, 481, 475

\bibitem[Wittenmyer et al.(2012a)]{HUAqr2} 
Wittenmyer, R.~A., Horner, J., Marshall, J.~P., Butters, O.~W., \& Tinney, C.~G.\ 2012a, \mnras, 419, 3258 

\bibitem[Wittenmyer et al.(2012b)]{HD142} 
Wittenmyer, R.~A., Horner, J., Tuomi, M. et~al., 2012b, \apj, 753, 169 

\bibitem[Wittenmyer et al.(2013)]{NSVS}
Wittenmyer, R.~A., Horner, J., \& Marshall, J.~P.\ 2013, \mnras, 431, 2150

\bibitem[Wittenmyer et al.(2014)]{Wittenmyer2014}
Wittenmyer, R.~A., Tan, X., Lee, M.~H. et al., 2014, ApJ, 780, 140

\bibitem[Yang et~al.(2010)]{YangEtAl2010} 
Yang, Y.-G., Li, H.-L., Dai, H.-F., Zhang, L.-Y., 2010, \aj, 140, 1687
\end{thebibliography}
\end{document}